# Efficient Packet Forwarding in Mesh Network

http://dx.doi.org/10.3991/ijim.v6i2.1991

Soumen Kanrar
Vehere Interactive Pvt Ltd, Calcutta, India

*Abstract*—Wireless mesh Network (WMN) is a multi hop low cost, with easy maintenance robust network providing reliable service coverage. WMNs consist of mesh routers and mesh clients. In this architecture, while static mesh routers form the wireless backbone, mesh clients access the network through mesh routers as well as directly meshing with each other. Different from traditional wireless networks, WMN is dynamically self-organized and self-configured. In other words, the nodes in the mesh network automatically establish and maintain network connectivity. Over the years researchers have worked, to reduce the redundancy in broadcasting packet in the mesh network in the wireless domain for providing reliable service coverage, the source node deserves to broadcast or flood the control packets. The redundant control packet consumes the bandwidth of the wireless medium and significantly reduces the average throughput and consequently reduces the overall system performance. In this paper I study the optimization problem in Wireless Mesh Networks. We have proposed a novel approach to reduce the broadcast redundant packet in the wireless mesh network. Also we have shown, a novel procedure to forward the control packet to the destination nodes and efficiently minimize the transmitted control packet in the wireless mesh cloud, that covers the domain.

## I. INTRODUCTION

A wireless mesh Network (WMN) is a communication network made up of radio nodes organized in a mesh topology. Wireless mesh networks often consist of mesh clients, mesh routers and gateways. The mesh clients are often laptops, cell phones and other wireless devices. The mesh routers forward traffic to and from the gateways which may but need not be connected to the internet. The coverage area of the radio nodes working as a single network is a mesh cloud. In this paper we have enhanced performance of the mesh cloud by optimized the control packet. The nodes operate in WMN not only as a host but also a router, forwarding packet s on behalf of other nodes that may not only be within direct wireless transmission range of their destinations. The mesh routers have minimal mobility and form the backbone of the WMNs. It provides network accesses for the both mesh and conventional client's.The Mesh clients can be either stationary or mobile and can form a client mesh network among themselves with mesh routers. A wireless mesh network can be seen as a special type of wireless ad hoc network. It's virtually appeared that all nodes in a wireless mesh network are immobile but practically it is not true. Researchers have started to revisit the protocol design of existing wireless networks specially of IEEE802.11 networks, Ad hoc network and wireless sensor networks.IEEE802.11, IEEE802.15 and IEEE802.16[ 3,8, 9] all have established sub working groups to focus on new standard for WMNs. WMN is a promising wireless technology for various application[7].Moreover the router bridge functionalities in mesh routers enables integration of WMNs with existing various wireless networks likes cellular Sensor networks, wireless–fidelity (Wi-Fi), WiMAX. The performance of any wireless system depends upon the selection of propagation mode for transferring control and data packets through multi hops [5]. In this paper I have focused how efficiently the minimum number of routers rebroadcast the hello or the beacon message. The router forwards the message that reaches all the nodes over the wireless mesh network. The packets are forwarded through the router nodes only. All the nodes apart from the router nodes are client nodes. Optimize flooding is done by minimizing the number of router nodes in the mess network in such a way that any client node can be reached through the least number of hops that passes through the least number of routers. The structure of the paper organized as, section I contains Introduction, Section II contains the system model for the wireless mesh cloud. Section III presents the packet transmitting algorithms. Section IV presents the simulation environment for the wireless mesh cloud. Section V presents the result and discussion. Section VI presents Conclusion and remarks.

### A. Network Architecture

The architecture of WMNs can be classified into three types, Infrastructure Backbone WMNs, Client WMNs and hybrid WMNs. In Infrastructure Backbone WMN architecture, mesh routers from an infrastructure for clients, as shown in figure 1 the dashed and solid lines indicate wireless and wired link, respectively.

The WMN infrastructure or backbone can be built using various types of radio technologies, in addition to the mostly used IEEE 802.11 technologies. The mesh router forms a mesh of self-configuring, self-healing links among themselves. Conventional clients with an Ethernet interface can be connected to mesh routers Via Ethernet links. For the conventional clients with the same radio technologies as mesh routers, they can directly communicate with mesh routers. If different radio technologies are used, clients must communicate with their base station s that has Ethernet connection s to mesh routers. In Client WMNs, the network provides peer to peer networks among client devices. In this type of architecture client nodes constitute the actual network to perform routing and configuration functionalities as well as providing end user application to the user. Mesh router is not required for client WMN. Client WMN is same like the general Ad hoc network but end user have extra functionalities of routing and self-configuration. The Hybrid WMN architecture is the combination of infrastructure and client meshing as shown figure 2. Mesh client can access the network through mesh routers as well as directly meshing with other mesh clients. The infrastructure provide connectivity to other networks likes Wi-Fi, WiMAX, cellular, the hybrid architecture have more advantage over WMN architecture. To improve the performance of WMN's architecture, the routing capabilities of client provide improved connectivity and coverage inside WMN's

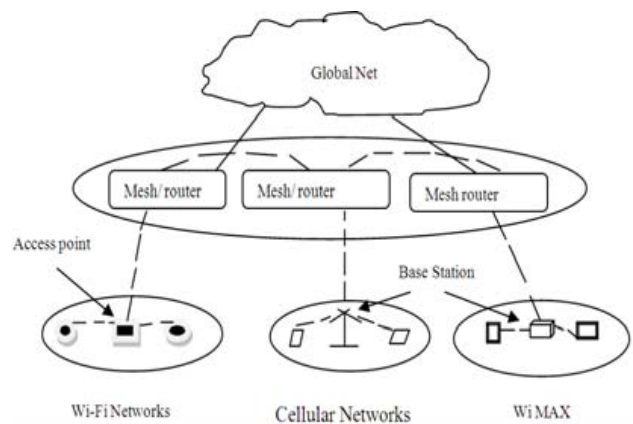

Figure 1. Infrastructure Backbone WMN

## II. SYSTEM MODEL

The system model develop to reduced the number of control packet and efficiently cover the mesh cloud. The router node can relay and retransmit the packets and the client node exchanges packets through the router nodes. We consider the packet exchange done by sharing of a common channel. Let $V = \{V_i \mid n \in \mathbb{N}\}$ is the set of all types of nodes in the wireless mesh network. Now $R \subset V$ be the sets of all routers in the wireless mesh network. In the proposed problem we consider The source { S} selects the nodes { $R_1$, $R_2$, $R_3$}. These are the router nodes from the set R to relay the message. When a flood message can be retransmitted from the nodes { $R_1$, $R_2$, $R_3$}. Any two hops neighbor must be covered by at least one node from the set R. The router forwards a flooding message with the following rule

i. The packet has not already been received.
ii. The router node is the last emitter.

that in the wireless mesh network, any 2 hops neighbor must be covered by at least one node $R_j \in R$. Any node $R_i \in R$ forwards a flooding packet with the condition that the packet has not already been received and that the node is the last emitter. To minimize the number of packets in the wireless domain and send the packet through least hops the below cardinality condition holds true,

1) The cardinality of the set (R) < cardinality of set (V)
2) Cardinality of set R < cardinality (V-R).

Now, Let S is the source node initiate a message such that message be reached every client node through the minimum number of hops comes from the source node S only for { $R_1$, $R_2$, $R_3$} then the message.

The message retransmitted follows this way. If the node A receives message from R2 for the first time then it will retransmit, if node A received from $R_1$ for the first time it won't re-transmit that packet. No client node will be missed to receive the message. Any node C (say) neighbor of the node A is a 2 hop neighbor of $R_1$. So any variable node which is at outer of the $2^{nd}$ circle be covered by the 2 hops of any node $R_i \in R$. We locally optimize the number of flooding packets. After that we proceed to optimize the flooding special packets by heuristic algorithms to cover the whole mesh network by transmitting minimum number of flooding packet. This is done to efficiently utilize the bandwidth of the wireless network and remarkably reduce the contention in the wireless domain. In this procedure we can reduce the race condition to occupy the sharable channel in the wireless domain. This procedure reduces the traffic load in the wireless domain. In wireless medium, random packet loss is a common issue due to random movement of the nodes in the wireless domain. The main disadvantage of the common proactive routing algorithms is the slow reaction on restructuring of the topology. The other is a respective amount of data used for maintenance. Here we have proposed a heuristic algorithm to reduce the traffic load in the wireless mesh domain.

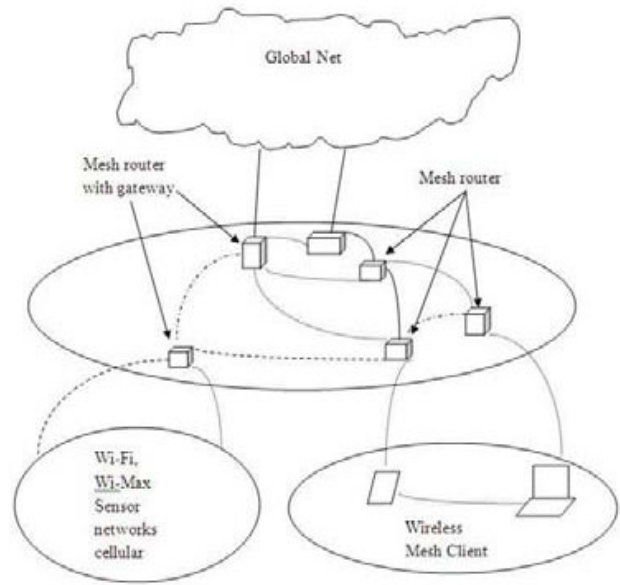

Figure 2. Hybrid WMN architecture

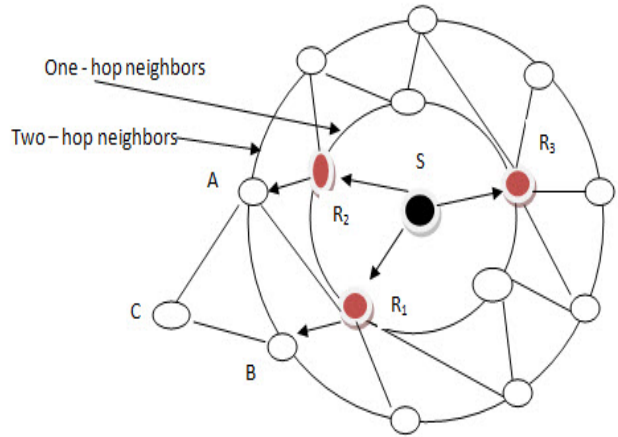

Figure 3. Architecture of the distributed nodes

## III. ALGORITHM

// let V and R are the set of nodes,

// $V = \bigcup_{i=1}^{n} v_i$ and initially $R = \varnothing$

Begin
For i=1 to n do

Select $v_i$ such that $v_i \in (V - R)$

For j=1 to n do

If ( $v_j$ is one covering of 2 hops neighbor of $v_i$ ) == True

then $R = R \cup \{v_i\}$

end for j
end for i

end begin // end algorithm

The time complexity of this heuristic algorithm is $O(n^2)$

## IV. SIMULATION ENVIRONMENT

OPNET was used to build the simulation model. All the operations have done by OPNET MODELER 16.0 PL1. For the simulation purpose, we have consider the number of grid uniformly distributed in the area of $500^2$ square meters. The sender node used the transmit power of 5m watts. The sender and receiver nodes used the channel capacity of 11 mbps in the mesh Cloud. We have used the interval between two emitting packets was 2 seconds. The mesh architecture configured dynamically by the topology control, in the interval of 5 seconds. The router in the mesh cloud holds the control packets for 6 seconds; this time is less than the topology control time. Since the node were dynamic in nature, the topology or a particular configuration stable for the 15 seconds. If the sender node sends any duplicate packet, that packet holds by the intermediate node for 30 seconds. The simulation runs for 300 seconds.

## V. RESULT AND DISCUSSIONS

Figure 4 represents the performance of the system model (1). It shows the traffic sent from the source node "s" according to the figure 3.
Initially the source node sends maximum control packet to cover the remote nodes with 2000 bits/ second.During the time 0 to 60 seconds the source node "s" send the packets within the range of (450 to 1000 bits per second). During the time interval 60+ seconds up to 300 seconds the simulation shows the source node send control packet within the range of (400 to 600 bits per second). The simulation shows clearly how the rate of transmission gradually decreases with respect to the increase of time. The simulation results (Fig. 4) clearly indicates how the traffic load reduces inside the mesh network.
The figure 5 represents the traffic passed through a router node. Initially the router $R_j \in R$ passes packets with the rate of 2200 bits per second. The extra 200 bits that is attached with the packet, which is the header of that router. During the time interval 10 to 300 seconds the router passes the packet with the rate of 2000 bits per second to 600 bits per second toward the next router.

Figure 6 represents the graphical view of the of the traffic received at the client node. This is more than 2 hops distance from the source node "s" via figure 3. Initially the client node receives packets with rate of 35000 bits per seconds. The client node received higher bits rate compare to the packet sent by the source node because of the header file that is attached with each control packet. During the time interval 60 seconds to 300 seconds the client node receives the control packet within the range of 15000 bits per second to 20,000 bits per seconds. The simulation figure clearly represents how the traffic load decreases inside the mesh network.

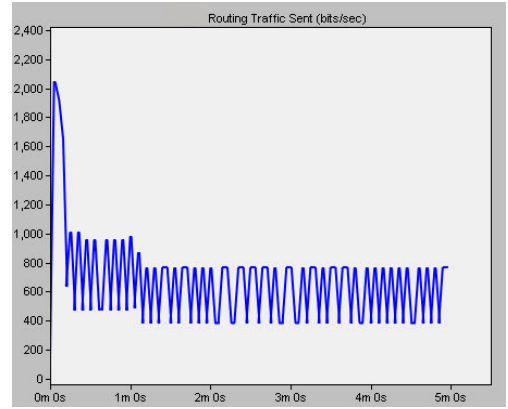

Figure 4. Traffic sent from the source node

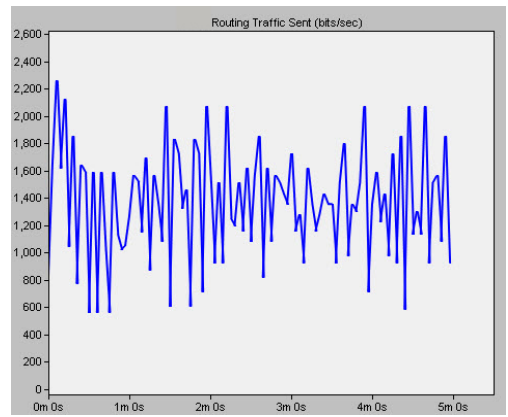

Figure 5. Traffic pattern passing through the router

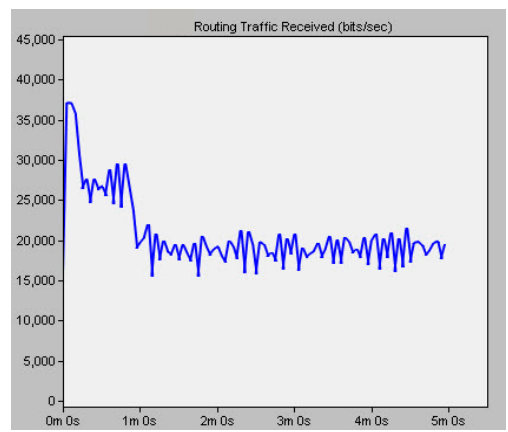

Figure 6. Traffic received at the client node

## VI. CONCLUSION REMARKS

In this work I have used a novel approach to reduce the traffic load in the wireless mesh network. To minimize the traffic load we try to minimize the number of the routers in the wireless mesh network. For finding the router nodes we have proposed a heuristic algorithm. The router nodes are eventually used to relay the control packet or rebroadcast the control packet. The time complexity of the heuristic algorithm is $O(n^2)$. In further work we can reduce the time complexity to nlog(n) as the packet is forwarded through the path which form a soft tree in the wireless mesh network.

## AUTHOR

**Soumen Kanrar** received the M.Tech. degree in computer science from Indian Institute of Technology Kharagpur India in 2000. Advanced Computer Programming RCC Calcutta India 1998. and MS degree in Applied Mathematics from Jadavpur University India in 1996. BS degree from Calcutta University India. Currently he is working as researcher at Vehere Interactive Calcutta India. Previously he had worked at King Saud University, Riyadh. Formally attached with the University Technology Malaysia. He is the member of IEEE. (email: Soumen.kanrar@veheretech.com).